\documentclass[prb,twocolumn,showpacs,preprintnumbers,amsmath,amssymb,float]{revtex4-1}
\usepackage{subfigure}
\usepackage{graphicx}
\usepackage{amsmath}
\usepackage{epstopdf}
\usepackage{amsbsy}
\usepackage{color}
\usepackage{dcolumn}
\usepackage{bm}

\begin{document}

\title{Impurity-induced antiferromagnetic order in Pauli-limited nodal superconductors: application to heavy fermion CeCoIn$_5$}

\author{Johannes H. J. Martiny$^1$, Maria N. Gastiasoro$^1$, I. Vekhter$^2$, and Brian M. Andersen$^1$}
\affiliation{$^1$Niels Bohr Institute, University of Copenhagen, Universitetsparken 5, DK-2100 Copenhagen,
Denmark\\
$^2$Department of Physics and Astronomy, Louisiana State University, Baton Rouge, Louisiana, 70803, USA}

\date{\today}

\begin{abstract}

We investigate the properties of the coexistence phase of itinerant antiferromagnetism and nodal $d$-wave superconductivity ($Q$-phase) discovered in heavy fermion CeCoIn$_5$ under applied magnetic field. We solve the minimal model that includes $d$-wave superconductivity and underlying magnetic correlations in real space to elucidate the structure of the $Q$-phase in the presence of an externally applied magnetic field.  We further focus on the role of magnetic impurities, and show that they nucleate the Q-phase at lower magnetic fields. Our most crucial finding is that, even at zero applied field, dilute magnetic impurities cooperate via RKKY-like exchange interactions to generate a long-range ordered coexistence state identical to the Q-phase. This result is in agreement with recent neutron scattering measurements [S. Raymond {\it et al.}, J. Phys. Soc. Jpn. {\bf 83}, 013707 (2014)].

\end{abstract}

\pacs{71.27.+a, 74.20.Rp, 74.70.Tx}

\maketitle

\section{Introduction}

Identifying new pathways towards realization of unusual ordered states is one of the main tasks of modern condensed matter physics~\cite{Keimer_review,PColeman}. Many correlated systems exhibit close competition of several electronic ordering phenomena~\cite{QSi,PColeman2,Stewart,Pfleiderer}. Since the energy difference between distinct ground states is small, pressure, chemical substitution, or external fields can be effectively used to tune the phase transitions, and switch the compounds between ordered states~\cite{Stewart,Pfleiderer,Stewart2}. One of the most studied examples is the competition and coexistence of superconductivity and itinerant magnetism~\cite{QSi,PColeman2,Stewart,Pfleiderer,Stewart2}, yet experiments still surprise with unexpected features and demand new theoretical ideas and approaches.

Layered heavy fermion CeCoIn$_5$ is one of those surprising materials. The nodal $d$-wave superconducting state, with the highest transition temperature, $T_c=2.3$K, among this class of compounds at ambient pressure, is formed before the coherence of the heavy electron state is reached, and the residual magnetic fluctuations have a strong effect on its properties~\cite{Thompson}. This compound is very close to the antiferromagnetic (AFM) transition as is evident from the transport and magnetic measurements~\cite{Bianchi1,JPaglione,Kohori,Onuki1,FRonning,CAlmasan}, neutron scattering results~\cite{Stock1}, as well as the appearance of the AFM phase upon alloying with Rh or Ir or doping with Cd ~\cite{GZheng,OnukiRh,Zapf,LPham,Park}. The behavior of the upper critical field, $H_{c2}(T)$, is consistent with strong Pauli-limiting of superconductivity~\cite{Bianchi4,Bianchi3}, i.e. the Zeeman splitting dominates over the orbital coupling in suppression of the superconductivity by an applied magnetic field. In part because of that the new thermodynamic phase (often referred to as Q-phase) discovered at low temperatures and high magnetic fields, $H\lesssim H_{c2}$, for the field parallel to the conducting planes, was suggested~\cite{BianchiFFLO} to be the long sought after Fulde-Ferrell-Larkin-Ovchinnikov (FFLO) state where the order parameter is spatially modulated at the wave vector proportional to the applied field.

Subsequent experiments~\cite{Vesna,Urbano,Kenzelmann1,Vesna2,Kenzelmann2,Forgan,Stock2} indicated, however, that this phase has long range incommensurate (IC) AFM order with a small ordered moment of order 0.4$\mu_B$, implying an itinerant nature of the magnetism. Surprisingly, and in contrast to other known cases, the AFM state only coexists with superconductivity but does not survive the destruction of pairing by the field, pressure, or doping. Several proposals argued that the FFLO modulation increases the density of states in the regions where the order parameter vanishes, $\Delta(\bm r)=0$, to above the normal state value, nucleating AFM order~\cite{Agterberg,Yanase1,Yanase2,Yanase3}.  Other theories rely on the microscopic conditions for the homogeneous coexistence of the two orders under a Zeeman field~\cite{Aperis1,Aperis2,Ikeda,kato11,Machida,kato12}. Physically, a very appealing scenario shows that improved  nesting of the near-nodal quasiparticle pockets created by the Zeeman field~\cite{kato11,Machida,kato12} yields AFM at the IC wave vector connecting the nodal points. The single lineshape of NMR signals~\cite{Vesna2,Kumagai} suggest uniform AFM order, and, together with the independence of the ordering wave vector on the field~\cite{Kenzelmann2}, seems to favor the latter scenario, although no consensus exists so far. Both types of theories are therefore tasked with explaining newly appearing experimental results within the same assumptions.

Our work is stimulated by a recent measurement that found a precise analog of the Q-phase in 5\% Nd-doped CeCoIn$_5$ already in the absence of the applied field inside the superconducting phase.\cite{Raymond14} The wave vector and even the magnitude of the ordered moment are identical to those in the Q-phase. While the authors of Ref.~\onlinecite{Raymond14} argued in favor of the nesting scenario as the most likely to produce the itinerant magnetic order, until now there have been no calculations analyzing its emergence in a disordered system. We consider this problem below.

One of our main findings is that a low concentration of magnetic impurities in nodal superconductors close to the AFM instability produces static itinerant magnetic order in the electron system. The key aspect of our calculation is accounting for the RKKY exchange interaction among the impurity spins by allowing their relaxation to a configuration that minimizes the free energy. We find that the induced magnetic order occurs at precisely the wave vector connecting the nodal points at the Fermi surface, in direct analogy with the Q-phase under applied field. To verify this we also show that, under the assumption of Pauli-limiting, dilute magnetic impurities extend the range in the $T$-$H$ plane where the Q-phase is stable. We emphasize, however, that our impurity calculation in the absence of the applied magnetic field does not rely on Pauli-limiting, and therefore our results are widely applicable to quasi-two-dimensional superconductors at the edge of magnetism, including organic compounds and other systems.

\section{Model and method}

\begin{figure}[b]
\subfigure{
\includegraphics[width = 8.cm]{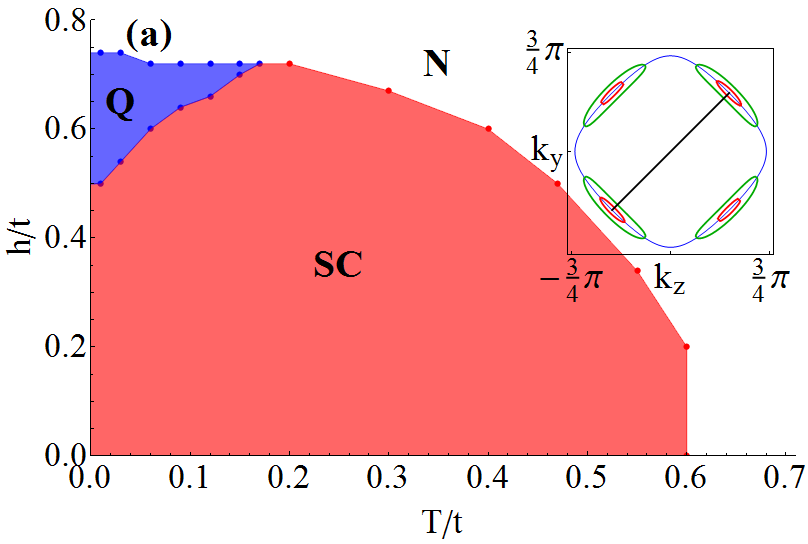}}\\
\hspace{-1.5em}
\subfigure{
\includegraphics[width=2.95cm]{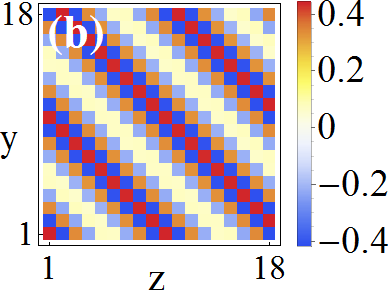}\hspace{-0.6em}}
\subfigure{
\includegraphics[width=2.95cm]{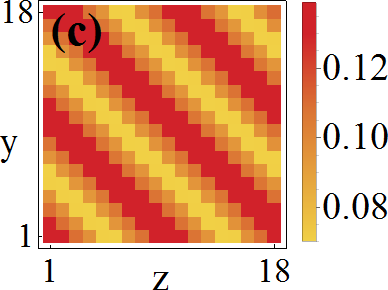}\hspace{-0.4em}}
\subfigure{
\includegraphics[width=2.95cm]{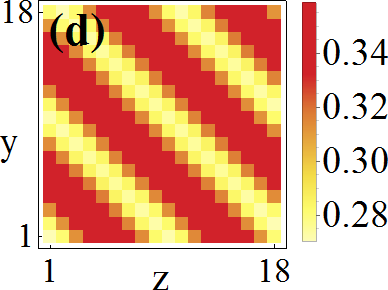}\hspace{-1.2em}}
\caption{(a) Phase diagram of the clean system including the Q-phase (blue region) at low $T$ and high magnetic fields $H$, the $d$-wave SC phase (red region), and the non-SC non-magnetic normal $N$ phase (white region).
The inset shows the normal state electron-like Fermi surface (blue curve) and two sets of contours of constant quasiparticle energy at $h=\{0.22,0.5\}$ (red, green) along with the nesting vector $\mathbf{Q}_+$ (black line) .
(b-d) Self-consistent magnetic $m^x(r)$, $m^z(r)$, and superconducting $\Delta(r)$ order parameters in the Q-phase, at $h= 0.7$ and $T=0.01$.
\label{fig:1}}
\end{figure}

CeCoIn$_5$ is a nodal superconductor with $d_{x^2-y^2}$-wave gap symmetry. In the presence of an external Zeeman magnetic field we describe it by the mean-field BCS Hamiltonian
\begin{align}
 \label{eq:H}
 \mathcal{H}_{sc}&= -t \sum_{\langle ij \rangle \sigma} c_{i \sigma}^\dagger c_{j \sigma} + \sum_{i\delta} (\Delta_{i\delta}\nonumber
c_{i\uparrow}^\dagger c_{i+\delta \downarrow}^\dagger + h.c.) \\
&~~~~ - \sum_{i\sigma }(\mu+\sigma h) c^\dagger_{i\sigma}c_{i\sigma}.
\end{align}
The first two terms describe the kinetic energy with nearest neighbor tight-binding hopping integral $t=1$, and the superconductivity with $d$-wave order parameter $\Delta_{i\delta} =  V\langle c_{i\uparrow}c_{i+\delta \downarrow}\rangle$, respectively. Here $\delta$ spans the vectors connecting nearest neighbors, and $V$ is the pairing potential.
The third term includes the chemical potential and the Zeeman splitting, $h= g\mu_B H$, associated with an in-plane (in the following defined as the $y-z$-plane) applied magnetic field $H$ (chosen along the in-plane $z$ axis).
The orbital part of the magnetic field is not included in the Hamiltonian due to the Pauli-limited nature of the system.
Finite values of $h$ result in small pockets of contours of constant quasi-particle energy located near the nodal regions in the superconducting state as shown in the inset of Fig.~\ref{fig:1}.
These field-generated pockets are perfectly nested at low fields, and introduce the possibility of itinerant magnetic order in the system.\cite{kato11,kato12}
Greater fields increase the effective bandwidth of the Bogoliubov quasiparticles, and preserve partial nesting, enabling the coexistence of AFM and $d$-wave superconductivity up to the first order transition in the paramagnetic normal state.
By fixing $\mu = -0.694$ we obtain that the nesting vectors are given by $\mathbf{Q}_{\pm}=(0,\pm 8/9\pi,8/9\pi)$.

Since the easy axis for the magnetization of these compounds is out-of-plane~\cite{Tayama,Curro} (here the $x$-axis), we include the magnetic interactions between the conduction electrons via an anisotropic exchange term of the form
\begin{align}
\mathcal{H}_{ex} &= - 2 \sum_{il}J_l s^l_i s_i^l,
\end{align}
which favors antiferromagnetic (AF) order perpendicular to the plane when $J_x>J_z,J_y$ and the exchange coupling exceeds a critical interaction strength $J_c$ in zero field.
The AF order parameter is $m^l(r_i) = \langle s_i^l\rangle$, where $l \in \{x,y,z\}$ with $ \langle s_i^l\rangle$ denoting the average spin density of the conduction electrons at site $i$. Finally, the system size $N\times N$ is chosen with $N$ being a multiple of 9 in order to accommodate an integer number of periodic modulations corresponding to $\mathbf{m}(\mathbf{r}_i)=\sum_{n=\pm} \mathbf{m}_n e^{i\mathbf{Q}_n\cdot \mathbf{r}_i}$.

By introducing a unitary Bogoliubov transformation $U$, the Hamiltonian $\mathcal{H}=\mathcal{H}_{sc}+\mathcal{H}_{ex}$ may be brought to diagonal form,
and we solve the resulting eigenvalue problem with the following self-consistency conditions
\begin{align}
n_{i\uparrow } &= \sum_{n=1}^{4N^2} |\alpha_{in}|^2 f(E_n), \\
n_{i\downarrow } &= \sum_{n=1}^{4N^2} |\omega_{in}|^2 f(E_n), \\
\Delta_{i\delta} &=  V\sum_{n=1}^{4N^2} \alpha_{in} \beta_{(i+\delta) n}^* f(-E_n),\\
\langle c_{i\uparrow}^\dagger c_{i\downarrow}\rangle &= \sum_{i=1}^{4N^2}\nu_{in} \beta_{in}^* f(-E_n).
\end{align}
Here $f(E)$ is the Fermi function, and $\alpha_{in}, \beta_{in}, \omega_{in}, \nu_{in}$ denote the $\{0,1,2,3\}\cdot N^2<i \leq \{1,2,3,4\}\cdot N^2$ interval of components of the $4N^2 \times 4N^2$ transformation matrix $U_{in}$, respectively.

\section{Q-phase in the magnetic field}

In this section we discuss the results obtained by solving the total Hamiltonian $\mathcal{H}=\mathcal{H}_{sc}+\mathcal{H}_{ex}$ under Zeeman field. We compare the results for the clean case with that for a single impurity, and show that the realm of the $Q$-phase is extended by magnetic disorder.

\subsection{The clean case}

In all of the following, we fix the $d$-wave pairing potential to $V =3.0$. The superconducting phase (in zero field) becomes unstable towards magnetic order at $J_c = 3.6$, and we consider a system close to this instability by fixing $J_x=3.5$, and further include a strong anisotropy of the form $J_z=J_y=J_x/14$. The exact values of $J_z$ and $J_y$ are not important for the main conclusions of this paper.
The obtained phase diagram for the homogeneous phase is shown in Fig.~\ref{fig:1}(a), and agrees well with the earlier results by Kato {\it et al.}\cite{kato11,kato12}
As seen from Fig.~\ref{fig:1}, the ground state at low $T$ is either a pure $d$-wave superconductor or a coexistence phase of IC magnetic order $\mathbf{m}(\mathbf{r}_i) = (m^x(\mathbf{r}_i),0,m^z(\mathbf{r}_i))$ and modulated $d$-wave superconducting order, constituting the so-called Q-phase.
The $m^x(\mathbf{r}_i)=m_+^xe^{i\mathbf{Q}_+\cdot \mathbf{r}_i}$ component of the single-Q phase is shown in Fig.~\ref{fig:1}(b).\cite{kato11}

\begin{figure}[t]
\begin{center}
\includegraphics[width=0.99\columnwidth]{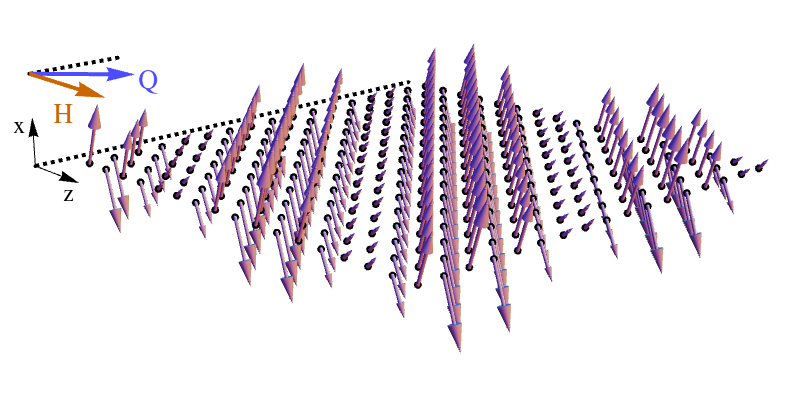}
\end{center}
\caption{3D plot of the real-space structure of the total magnetization of the single-Q phase.
The orange and blue arrows indicate the direction of the applied field $H$ and the periodicity of the SDW $\mathbf{Q}_{+}$, respectively.}
\label{fig:2}
\end{figure}

The present unrestricted self-consistent Hartree-Fock study allows to go beyond the uniform coexistence explored in Refs. \onlinecite{kato11,kato12}, and obtain the real-space modulated structures in the magnetization, $m^z(\mathbf{r}_i)=m_0^z+m_+^ze^{i2\mathbf{Q}_+\cdot \mathbf{r}_i}$, and the SC order parameter $\Delta(\mathbf{r}_{i})=\Delta_0+\Delta_+ e^{i2\mathbf{Q}_+\cdot \mathbf{r}_i}$,
defined by
\begin{align}
\Delta(\mathbf{r}_{i}) &=\frac{1}{4} \sum_{\delta} |\Delta_{i}^s|=\frac{1}{8} \sum_{\delta} |\langle c_{i\uparrow}c_{(i+\delta) \downarrow } \rangle - \langle c_{i\downarrow}c_{(i+\delta) \uparrow } \rangle|.
\end{align}
The results are shown in Figs.~\ref{fig:1}(c,d). As seen from Fig.~\ref{fig:1}(d), the superconducting order parameter is suppressed in regions of maximum magnetization in agreement with standard competitive behavior between these two order parameters. The magnetic structure of the single-Q phase is also displayed in Fig.~\ref{fig:2}, which explicitly shows the spatial profile of the orientation of the moments obtained by combining Figs.~\ref{fig:1}(b) and \ref{fig:1}(c).

Note that we have not considered here the possible double-Q phase that may exist in a narrow range of fields and temperatures~\cite{kato11,kato12}. Recent arguments show that spin-orbit interaction stabilizes a single-Q phase cross the phase diagram~\cite{Mineev}, and can explain the observed domain switching under field rotation in the basal plane~\cite{Kenzelmann3}.

\subsection{Single magnetic impurity under applied field}

Having discussed the clean Q-phase we now turn to the inclusion of a single magnetic impurity placed at site $i^*$.
We investigate two distinct cases; 1) the impurity fully aligned with the field along the $z$-axis (Z-impurity), and 2) the impurity sharing the same easy axis as the conduction electrons of the Q-phase, which is defined as the out-of-plane $x$-axis shown in Fig.~\ref{fig:2} (X-impurity). The associated impurity part of the Hamiltonian is given by
\begin{align}
\mathcal{H}_{imp} &=- \sum_{l} j_l S^l_{i^*} s_{i^* }^l
\end{align}
where $S^l$ represents the local impurity spin interacting with the spin of the conduction electrons $s^l$, and $l=x,z$ for an X- and Z-impurity, respectively.
Of course, the physical situation may also arise that an $x$-aligned impurity rotates (aligns) with the applied field, leading to a mixture of X- and Z-terms. This situation can also be understood from the results presented below. In this paper, we follow the standard approach by Shiba~\cite{shiba} and model the magnetic impurity spin in the classical approximation corresponding to the limit $S^l\rightarrow \infty$, $j_l\rightarrow 0$ with $j_l S^l$ finite.

\begin{figure}[b]
\subfigure{
\includegraphics[width = 4.05cm]{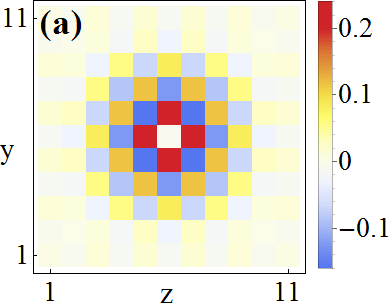}}
\subfigure{
\includegraphics[width = 4.2cm]{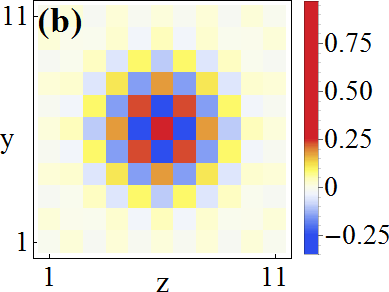}} \\

\hspace{-0.32cm}
\subfigure{
\includegraphics[width = 4.3cm]{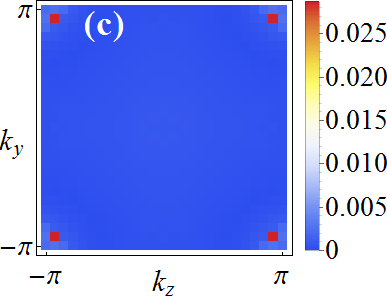}}
\hspace{-0.22cm}
\subfigure{
\includegraphics[width = 4.07cm]{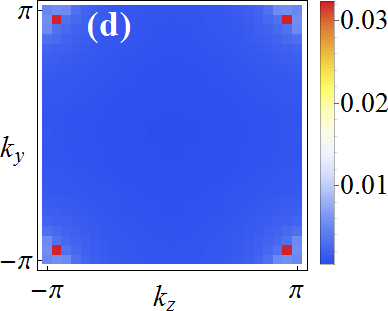}}\\

\hspace{-0.3cm}
\subfigure{
\includegraphics[width = 3.8 cm]{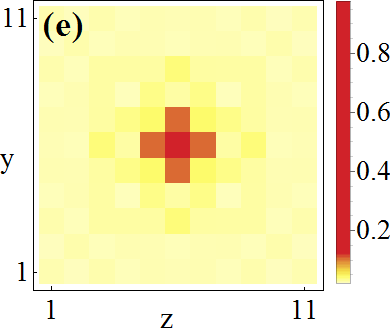}}
\hspace{0.3cm}
\subfigure{
\includegraphics[width = 4.cm]{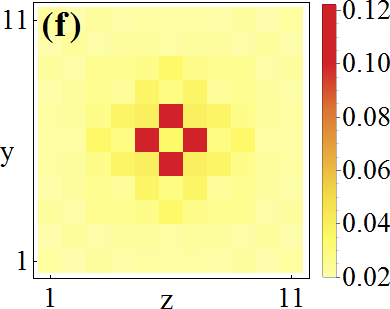}}
\caption{Zoom of the induced magnetization of a single $Z$-impurity (a,c,e), and a single $X$-impurity (b,d,f) in an external magnetic field along the $z$ axis of sub-critical amplitude $h=0.45$. $j_l S_{i^*}^{l}=20$ in both cases.
We show in (a,b) the $m_x$ component, (c,d) the Fourier transform of this component, and (f,e) the $m_z$ component along the applied field. For all cases here; $T= 0.01$ and the system size is $27 \times 27$.}
 \label{fig:3}
\end{figure}

\begin{figure}[t]
\includegraphics[width = 8.2cm]{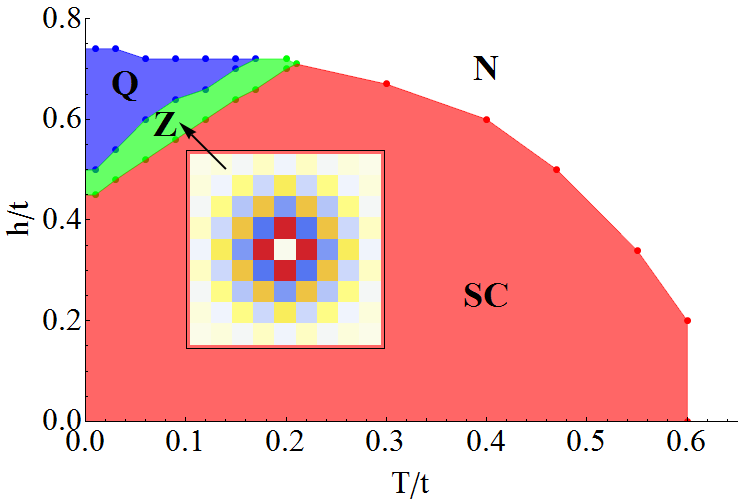}
\caption{Phase diagram of the inhomogeneous system displaying an extended Q-phase where isolated magnetic Z-impurities (oriented along the $z$ axis) locally nucleate magnetization with the same local structure as in the magnetically long-range ordered Q-phase, i.e. along the $x$ axis.}
\label{fig:4}
\end{figure}

A single Z-impurity in a magnetic field below the onset of the Q-phase ($h<h^*=0.5$, $T=0$), can form a local pocket of perpendicular $m_x$ magnetization if the coupling to the conduction electrons is large enough ($j_z S_{i^*}^z >1.7$).
We show in Figs.~\ref{fig:3}(a) and (e) the induced magnetization perpendicular to the field ($x$-axis) and along the field ($z$-axis).
The Fourier transform of $m_x(\mathbf{r}_{i})$ exhibits peaks corresponding to the nesting vectors  $\pm\mathbf{Q}_{\pm} =(\pm 8/9 \pi, \pm 8/9\pi)$ for the $m^x(\mathbf{q})$ component as seen from Fig.~\ref{fig:3}(c).
These results can be understood by following the evolution of impurity resonant states close to the Fermi energy as a function of the coupling $j_l$ and the applied field $h$.
The local density of states at the impurity site and nearest neighbors is enhanced at low energies by the resonant state formation, which allows for a local Stoner instability criterion to be fulfilled.
As the applied field $h$ decreases, the LDOS enhancement is similarly reduced for all couplings $j_l$, until at some critical $h_{c}\sim 0.45$ the local Stoner criterion can no longer be satisfied. Notably, this disorder-phase effectively extends the Q-phase in both field ($h_{c}<h<h^*$) and temperature.
This result is displayed explicitly in Fig.~\ref{fig:4} where a new Z-phase induced by Z-impurities indicates the region of stability of the local impurity-induced magnetic Q-structured spin polarization along the $x$-axis (inset).
In this way, in the dilute limit, disorder due to magnetic impurities is not detrimental to the Q-phase, but rather acts to expand its local manifestations in the phase diagram. Local probes, such as NMR or STM, should see evidence for these locally nucleated magnetic puddles, which may even, if the concentration of magnetic impurities is sufficiently large, couple and generate a long-range ordered phase as discussed further below.

For completeness we show also in Fig.~\ref{fig:3}(b,d,f) the magnetization components for an X-impurity in a finite field.
This case is relevant to materials where the external field is too weak to align the moments of the magnetic impurities along the external field direction.
As seen from Fig.~\ref{fig:3}(b), also in this case a significant impurity-induced $m^x$ magnetization is generated, along with a minor $m^z$ component [Fig.~\ref{fig:3}(f)] which decays as the field is decreased.
The Fourier transform of the $m^x$ component is again sharply peaked at Q as shown in Fig.~\ref{fig:3}(d).

We note that the properties of impurity-induced magnetization in $d$-wave superconductors have been discussed quite extensively in the context of the high-T$_c$ cuprates.\cite{Tsuchiura,Wang,zhu02,Chen,Kontani,Harter,Andersen07,Andersen10,Christensen} There, however, the main focus has been on non-magnetic disorder arising e.g. from the dopants, and the explanation of how electronic correlations may lead to local nucleated magnetic order and spin-glass phases. This situation is different from the present case, where magnetic Ising-like impurities directly couple to the magnetic susceptibility of the host material.

\section{Cooperative impurity-induced Q-phase at zero magnetic field}

\begin{figure}[t]
\subfigure{
\includegraphics[width = 4.1 cm]{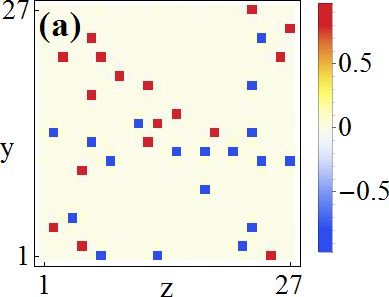}} \\
\subfigure{
\includegraphics[width = 4.1cm]{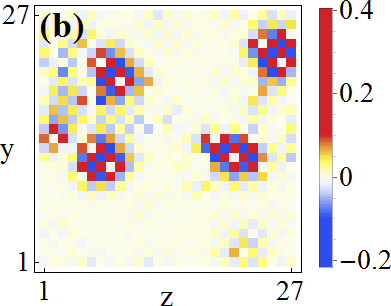}}
\subfigure{
\includegraphics[width = 4.2cm]{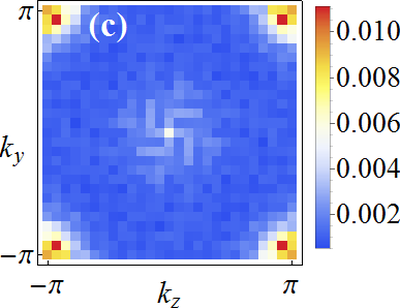}}
\caption{(a) Real space $m^z(r)$ component for a sample system of $5\%$ Z-impurities ($j_z S^z_{i^*}=20$) in the absence of an external magnetic field ($H=0$). The corresponding FT is simply noise. (b) Induced magnetization $m^x(r)$ for the same system. (c) Averaged $|m^x(\mathbf{q})|$ including three of these Z-impurity configurations.}
\label{fig:6}
\end{figure}

\begin{figure}[b]
\subfigure{
\includegraphics[width = 4.15cm]{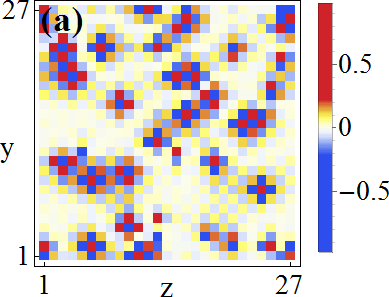}}
\hspace{-0.3cm}
\subfigure{
\includegraphics[width = 4.15cm]{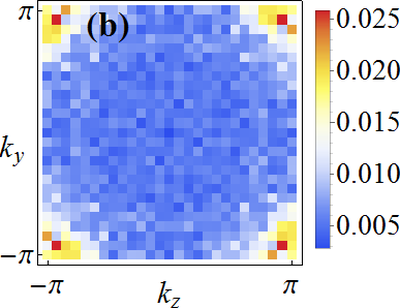}}
\caption{ (a) Real space $m^x(r)$ component induced by a relaxed system of $5\%$ X-impurities ($j_x S^x_{i^*}=20$) in the absence of an external magnetic field ($H=0$).
(b) Averaged $|m^x(\mathbf{q})|$ including three different configurations, exhibiting sharp peaks at $\mathbf{Q}_{\pm}$ arising from cooperative exchange.}
\label{fig:5}
\end{figure}

The single impurity ordering discussed in the previous section leads to the hypothesis that the experimental results obtained for Nd-doped CeCoIn$_5$ by Ref.~\onlinecite{Raymond14} at zero field ($H=0$) may be explained by a cooperative impurity scenario. In the following we study two distinct cases 1) a concentration of magnetic Z-impurities cooperating to induce a long-range ordered $m^x$ component (in analogy with the homogeneous case with a field along the $z$-axis), or 2) a concentration of X-impurities each of which nucleate local $m^x$ order which couple and thereby generate a long-range ordered state with sharp peaks at $\mathbf{Q}$. Naturally, the orientation of the impurity moments may also lead to a combination of 1) and 2). To investigate these possibilities we study systems doped with $5\%$ magnetic impurities with moments aligned along either the $x$- or the $z$-axis at zero magnetic field ($H=0$).
The impurities are randomly positioned within the lattice, subject only to the criterion that they not be near- or next-nearest neighbors. The moment of each (Ising) impurity is initially assigned a random direction along the anisotropy axis, but the system is subsequently allowed to "relax" by minimization of the total free energy $\mathcal{F} = \mathcal{U}-T\mathcal{S}$, where  the internal energy $\mathcal{U}$ and the entropy $\mathcal{S}$ are given by
\begin{align}
\mathcal{U} &= \langle \mathcal{H}^{MF} \rangle = \langle \mathcal{H}_{sc} \rangle + \langle \mathcal{H}_{ex}\rangle+ \langle \mathcal{H}_{imp}\rangle,\\
\mathcal{S}&=-k_B \sum_n \left[ f(E_n)\ln f(E_n) +  f(-E_n)\ln f(-E_n)\right].
\end{align}
More specifically, we impose impurity spin flips and thereby allow the system to optimize its magnetization configuration in order to minimize $\mathcal{F}$ in a standard Monte Carlo-like fashion.

Discussing first scenario 1), we show in Fig.~\ref{fig:6} the results of the magnetization generated by a collection of Z-impurities. Specifically, Fig.~\ref{fig:6}(a) and \ref{fig:6}(b) display the optimized results for $m^z(r)$ and $m^x(r)$, respectively.
As seen, the Z-impurities cooperatively induce long-range ordered magnetization only along the $x$ axis.
In this scenario the configuration-averaged Fourier map of $|m^x(q)|$ exhibits sharp peaks at $\mathbf{Q_{\pm}}$ as shown in Fig.~\ref{fig:6}(c).

\begin{figure}[t]
\subfigure{
\includegraphics[width = 4.1cm]{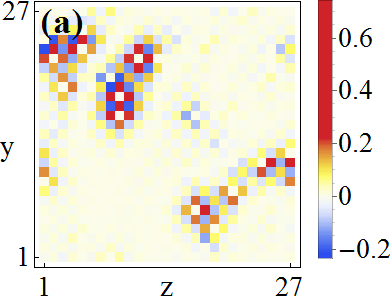}}
\subfigure{
\includegraphics[width = 4.2cm]{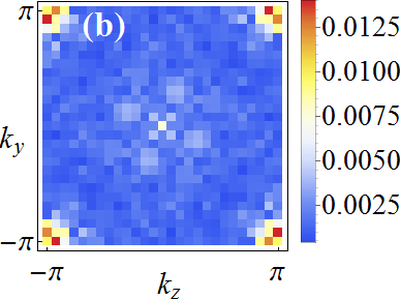}}
\caption{(a) Magnetization in real space for a set of unrelaxed Z-impurites. (b) The Fourier transform of (a) clearly showing the absent sharp $\mathbf{Q}$ peaks of the corresponding relaxed system in Fig.~\ref{fig:6}(c).}
\label{fig:7}
\end{figure}

Turning next to the second scenario, 2), we show in Fig.~\ref{fig:5}(a) the final "annealed" magnetization resulting from a given configuration of X-impurities at $H=0$.
As seen from this real space map an extended magnetic structure is clearly generated by the nucleated magnetic order around each X-impurity.
The configuration-averaged Fourier map $|m^x(\mathbf{q})|$ is shown in Fig.~\ref{fig:5}(b), displaying the same momentum structure as the clean Q-phase with prominent sharp peaks only at $\mathbf{Q_{\pm}}$.
This indicates that both cases discussed here are relevant, i.e. the experimentally obtained peaks can be explained both by $5\%$ doping with the impurities acting as local effective fields in the plane (Z-impurities) or with the moments along the conduction electron spin quantization axis (X-impurities). This cooperative RKKY exchange effect between the disorder is very similar to a recent theoretical explanation of the emergence of long-range magnetic order by substitution of Mn ions in BaFe$_2$As$_2$.\cite{gastiasoro14,kim,inosov13}

In order to explicitly demonstrate the cooperative impurity effects of the final magnetization, we show a result for a non-optimized case.
Focusing on the case of Z-impurities, Fig.~\ref{fig:7}(a) shows a snapshot of the magnetization before the moments are allowed to flip and lower the free energy.
The resulting magnetic order nucleated around the impurities shows broadened peaks in momentum space as evident in Fig.~\ref{fig:7}(b), where the expected sharp peaks at $\mathbf{Q}_{\pm}$ are notably absent. This is caused by the random directions of the impurity spins which prohibits the system from forming long-range order.

\section{Conclusions}

In summary, we studied a $d$-wave superconductor that is close to the AFM state by solving the self-consistent Bogoliubov-de Gennes equations in real space. For the impurity-free case in the presence of an in-plane (Zeeman) magnetic field, we find a region of coexistence of magnetism and superconductivity at low temperatures and high fields, in full agreement with previous studies of this, so called Q-phase.\cite{kato11,kato12} In going beyond previous approaches, we also showed that both the AFM magnetization and the superconducting order are spatially modulated in this phase.

In the bulk of the paper we investigated the influence of dilute magnetic impurities coupled to conduction electrons by an anisotropic exchange interaction in such superconductors. We showed how isolated magnetic impurities locally nucleate magnetic order consistent with the Q-phase even at fields below the onset of the global AFM-SC coexistence. Local magnetic probes, such as NMR, should be able to identify the signatures of such local Q-order.

Finally, we showed that even {\em at zero magnetic field} for a finite concentration of magnetic impurities, the induced RKKY-like interaction tends to align the impurity moments and generate a long-range ordered magnetic Q-phase. This result is in agreement with recent experimental findings on Nd-doped CeCoIn$_5$.\cite{Raymond14} Since this calculation only relies on the proximity in energy of the bulk AFM phase, and the exchange coupling of the classical impurity spin to the conduction electrons, we expect that very similar physics should be at play in, for example, two-dimensional organic superconductors.

We considered a purely magnetic impurity potential. Inclusion of the potential scattering may further suppress the transition temperature and smear the signatures of the $Q$-phase, but we leave a detailed study of this problem for future work. 

\section{Acknowledgements}

B.M.A. and M.N.G. acknowledge support from a Lundbeckfond fellowship (grant A9318). I. V. is supported in part by NSF Grant DMR-1105339.

\end{document}